\documentclass[twocolumn,showpacs,amsmath,prb]{revtex4}
\usepackage{graphicx}
\usepackage{dcolumn}
\usepackage{bm}

\begin{document}

\title{Nano granular metallic Fe - oxygen deficient TiO$_{2-\delta}$ composite films: 
A room temperature, highly carrier polarized magnetic semiconductor}

\author{S. D. Yoon}
\altaffiliation[Corresponding author e-mail: syoon@ece.neu.edu ]{}
\author{C. Vittoria}
\email{vittoria@lepton.neu.edu}
\author{V. G. Harris}
\email{harris@ece.neu.edu}
\affiliation{Center for Microwave Magnetic Materials 
and Integrated Circuits, Department of Electrical and 
Computer Engineering, Northeastern University, 
Boston, MA. 02115 USA}

\author{A. Widom}
\email{allan.widom@gmail.com}
\affiliation{Department of Physics, Northeastern University, 
Boston, MA. 02115 USA.}

\author{K. E. Miller}
\author{M. E. McHenry}
\affiliation{Department of Material Science and Engineering, 
Carnegie Mellon University, Pittsburg, PA. 15213, USA.}

\begin{abstract}
Nano granular metallic iron (Fe) and titanium dioxide (TiO$_{2-\delta}$) were co-deposited 
on (100) lanthanum aluminate (LaAlO$_3$) substrates in a low oxygen chamber pressure 
using a pulsed laser ablation deposition (PLD) technique. The co-deposition of Fe 
and TiO$_2$ resulted in $\approx$ 10 nm metallic Fe spherical grains suspended within 
a TiO$_{2-\delta}$ matrix. The films show ferromagnetic behavior with a saturation magnetization 
of 3100 Gauss at room temperature. Our estimate of the saturation magnetization based on the size 
and distribution of the Fe spheres agreed well with the measured value. 
The film composite structure was characterized as p-type magnetic semiconductor at 300 K 
with a carrier density of the order of $ 10^{22} /{\rm cm^3}$. The hole carriers 
were excited at the interface between the nano granular Fe and TiO$_{2-\delta}$ matrix 
similar to holes excited in the metal/n-type semiconductor interface commonly observed 
in Metal-Oxide-Semiconductor (MOS) devices. From the large anomalous Hall effect directly 
observed in these films it follows that the holes at the interface were strongly
spin polarized. Structure and magneto transport properties suggested that these PLD films have 
potential nano spintronics applications.
\end{abstract}


\maketitle
\section{introduction}
The search for semiconductors exhibiting magnetism at room temperature has been long and 
unyielding. However, recently much progress has been made toward this 
goal\cite{Ohno:1998,Ueda:2001,Masumoto:2001,Pearton:2003,Chambers:2003,Zutic:2004,Coey:2005}. 
By doping a host semiconductor material with transition metal ferromagnetic atoms, 
dilute ferromagnetic semiconductors have been produced with Curie temperatures ($T_c$) 
as high as 100 K\cite{Ohno:1998}. Hall effect measurements below $T_c$ showed evidence 
for carriers being spin polarized raising hopes for spintronics applications. 
Specifically, metallic manganese (Mn) was doped into gallium arsenide (GaAs) 
whereby the carriers were strongly spin polarized\cite{Ohno:1998}.

We have previously reported the magnetic and dc-transport properties of magnetic semiconductor 
films of TiO$_{2-\delta}$, where $\delta$ indicates the degree of oxygen deficiency or defects 
in the film\cite{Hong:2006,Yoon:2006,Yoon1:2007}. The Curie temperature, $T_c \approx$ 880 K, 
was well above room temperature with a saturation magnetization, $4\pi M_S \approx $ 400 Gauss. 
Titanium dioxide, TiO$_2$, is a well known wideband gap oxide semiconductor, belonging 
to the group IV-VI semiconductors, described in terms of an ionic model of 
Ti$^{4+}$ and O$^{2-}$\cite{Earle:1942,Breckenridge:1953,Daude:1977,Pascual:1978,Tang:1993}. 
Its intriguing dielectric properties allow its use as a gate insulator material 
in the Field-Effect-Transistor (FET)\cite{Cambell:1999}. Also, TiO$_2$ is characterized to be 
an n-type semiconductor with an energy gap varying in the range 3 volt $<\ \frac{\Delta}{e}\ <$ 9 volt 
depending on sample preparation\cite{Earle:1942,Breckenridge:1953,Daude:1977,Pascual:1978,Tang:1993}. 
Films of TiO$_{2-\delta}$ on substrates of (100) lanthanum aluminate (i.e. LaAlO$_3$) were deposited 
by a pulsed laser ablation deposition (PLD) technique at various oxygen chamber pressures ranging 
from 0.3 to 400 mtorr. The origin of the presence of Ti$^{2+}$ and Ti$^{3+}$ (as well as Ti$^{4+}$) 
ions was postulated as a result of the low oxygen chamber pressure during the films growth\cite{Yoon:2006}. 
Oxygen defects gave rise to valence states of Ti$^{2+}$ and Ti$^{3+}$ (in the background of Ti$^{4+}$) 
whereby double exchange between these sites dominated. The same carriers involved in double exchange 
also gave rise to impurity donor levels accounting for the transport properties of the film. 
The dilute number of carriers were polarized in external applied magnetic field, 
yet the magnetic moment was still rather small\cite{Yoon1:2007}. For example, 
normal Hall resistivity was measured to be much bigger than the anomalous Hall resistivity\cite{Yoon1:2007}.
In order to increase the anomalous contribution to the Hall effect and thereby increase 
the number of polarized carriers, we have fabricated a nano-granular (NG) metallic iron (Fe) 
in semiconducting TiO$_{2-\delta}$ matrix. The intent was to introduce a substantial 
magnetization component internally to the semiconductor TiO$_{2-\delta}$.

The basic difference between our film composite and the previously reported magnetic semiconductors 
by others is that in our films the NG of metallic Fe co-exist in a metallic state, 
whereas magnetic semiconductors prepared by others the transition metal co-exists as metal oxides 
and often oxide clusters\cite{Pearton:2003,Chambers:2003,Zutic:2004,Coey:2005}. This major difference is 
important in terms of the magnetic and transport properties of the magnetic semiconductor presently produced 
by us and that of others\cite{Pearton:2003,Chambers:2003,Zutic:2004,Coey:2005}. For example, NG metallic Fe 
contains a significant higher moment and Curie temperature 
than any other transition metal oxides. Also, the presence of NG metallic Fe allows 
for the creation of a reservoir of conduction electrons in the conduction band and, therefore, 
holes in the TiO$_{2-\delta}$  matrix. As electrons from the conduction band of TiO$_{2-\delta}$ 
are thermally introduced into the metallic Fe conduction band, holes are created in TiO$_{2-\delta}$ 
much like in junctions of NG metal/semiconductor interfaces or in Metal Oxide Semiconductor (MOS) 
devices\cite{Streetman:1990}. Conduction of holes occurs in the TiO$_{2-\delta}$ host. 
This mechanism gives rise to lower resistivity at high temperature in contrast to the pure TiO$_{2-\delta}$ 
reported earlier\cite{Yoon:2006,Yoon1:2007} where the carriers were only electrons. No such mechanism 
is possible in magnetic semiconductors doped with transition metal oxides. We report here that 
in our composite films of NG metallic Fe in anatase TiO$_{2-\delta}$ the magnetization, $4 \pi M_S \approx$ 
3,100 Gauss (also $\approx 0.4 \mu_B$/Fe), at room temperature, $T_c$ above 800 K, the carriers 
are strongly spin polarized and the room temperature resistivity is lowered by a factor of $\approx 10^{-3}$ 
relative to films of the undoped TiO$_{2-\delta}$ semiconductors previously 
produced\cite{Hong:2006,Yoon:2006,Yoon1:2007}. Experimental results, discussions and concluding 
remarks follow.

\section{Experimental Results and Analysis}
Thin films consisting of nano granular (NG) metallic Fe and TiO$_{2-\delta}$ were deposited 
by a pulsed laser ablation deposition (PLD) technique from binary targets of TiO$_2$ and metallic
Fe on (100) LaAlO$_3$ substrates. Targets of TiO$_2$ and Fe were mounted on a target rotator driven 
by a servomotor and synchronized with the trigger of the pulsed excimer laser $\lambda =$ 248 nm. 
In each deposition cycle, the ratio of laser pulses incident upon the TiO$_2$ target to those 
upon the Fe target was 6:1, this technique denoted as alternating targets-pulsed 
laser ablation deposition (AT-PLD) technique. The substrate temperature, laser energy density, 
and pulse repetition rate were maintained at 700 $^{\rm o}{\rm C}, \approx 8.9 \ {\rm J /cm^2}$, 
and 1 Hz, respectively. The deposition was carried out in a pure oxygen background of around $10^{-5}$ torr 
in order to induce defects in the TiO$_2$ host. There were a total of 4,206 laser pulses (3,606 pulses 
on TiO$_2$ target and 600 pulses on Fe target) for each film resulting in a thickness of 
approximately 200 nm as measured by a Dek-Tek step profilometer. 
\begin{figure}[tp]
\centering
\includegraphics[width=0.42\textwidth]{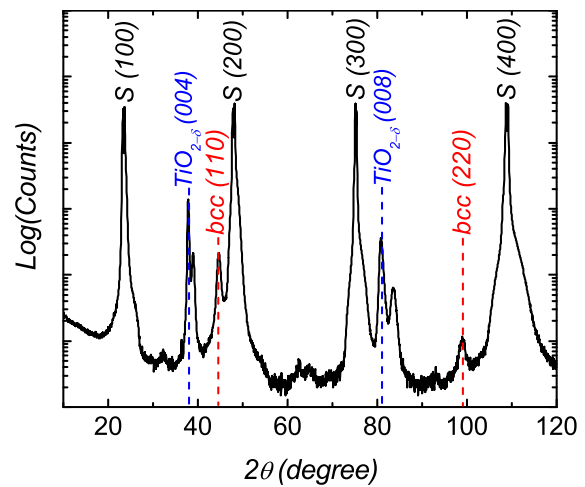}
\caption{X-ray diffraction pattern shows anatase of {\it (001)} TiO$_{2-\delta}$ and {\it (110)} 
bcc phase.}
\label{XRD} 
\end{figure}
\begin{figure}[bp]
\centering
\includegraphics[width=0.352\textwidth]{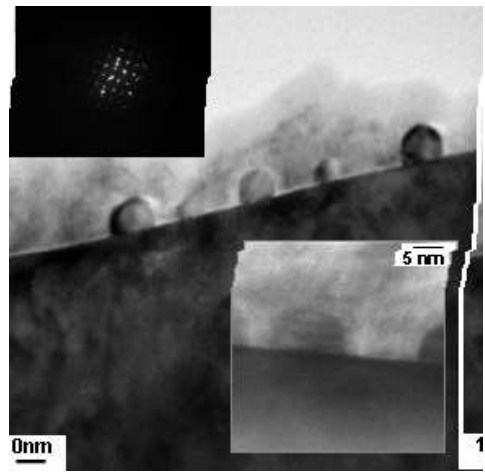}
\caption{TEM images for a representative film of nano granular Fe in TiO$_{2-\delta}$.}
\label{TEM} 
\end{figure} 

Crystallographic and micrographic properties of the AT-PLD films were measured using x-ray diffractometry 
(XRD) and transmission electron microscopy (TEM). Results indicate phases of anatase TiO$_2$, 
metallic body centered cubic (bcc) Fe. {\it (001)} plane of anatase TiO$_{2-\delta}$ and {\it (110)} plane 
of bcc Fe phase are clearly exhibited in the xrd pattern\cite{NIST:1969,Howard:1991,Swanson:1955} 
shown in FIG.\ref{XRD}, whereas peak appeared at $2\theta\  = \ 32.20 \ ^{\rm o}$ could not be readily indexed. 
From our speculation, the peak may be originated by possible presence of iron oxide 
(FeO or Fe$_2$O$_3$), or ilmenite (FeTiO$_3$) phase.

In FIG.\ref{TEM}, reflection electron diffraction (see upper inset) supports the existence 
of the epitaxial TiO$_{2-\delta}$ film host while the TEM image (and lower inset) 
reveal the presence of nano granular (NG) metallic Fe particles suspended within the TiO$_{2-\delta}$ host. 
A JOEL 2000EX high resolution transmission electron microscope operating at 200 keV and 400 keV was used 
in the analysis. A HRTEM image shown in FIG.\ref{TEM} illustrates that NG metallic Fe spheres were 
formed at the interface between TiO$_{2-\delta}$ layers and LaAlO$_3$ substrate. Average diameter of 
NG spheres was measured to be in the order of $\approx$ 10 to 15 nm as shown in the TEM image. 
The image suggests that the AT-PLD process may lead implantation of NG metallic Fe spheres 
into host TiO$_{2-\delta}$. Average atomic ratio of Fe/Ti ions of the AT-PLD films 
was obtained by energy dispersive x-ray spectroscopy (EDXS) within an ultra high resolution 
scanning electron microscope (UHR-SEM) column of Hitachi S-4800 
that $\approx \ 6 \ \pm \ 0.5 \%$ of Fe are possibly presented in the $1 \ \times \ 1 \ {\rm cm^2}$ 
area of the films.  
\begin{figure}[tp]
\centering
\includegraphics[width=0.43\textwidth]{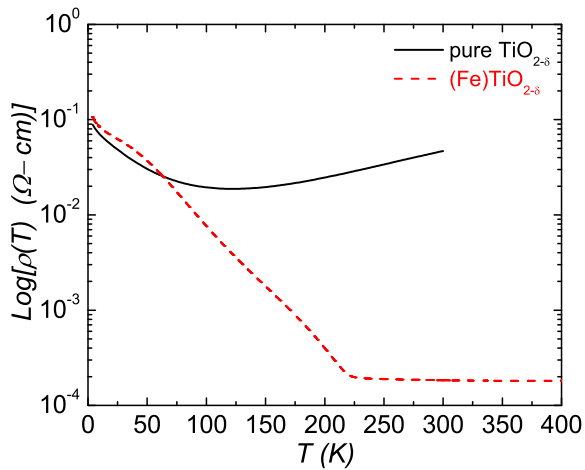}
\caption{Resistivity, $\rho$ as a function of temperature, $T$, for the nano granular Fe 
in TiO$_{2-\delta}$ film (dashed line) and the pure TiO$_{2-\delta}$ (solid line) 
from the reference \cite{Yoon1:2007}.}
\label{Resistivity} 
\end{figure} 

For steady currents and with the magnetic intensity {\bf H} directed normal to the film, 
the resistance matrix {\bf R} in the plane of the film may be written as\cite{Landau:1984} 
\begin{equation}
{\sf R}=
\begin{pmatrix}
R_{xx} & R_{xy} \\ 
R_{yx} & R_{yy}
\end{pmatrix}
=\frac{1}{t}
\begin{pmatrix}
\rho & -\rho_H \\ 
\rho_H & \rho
\end{pmatrix}
\label{Matrix}
\end{equation}
wherein $\rho$ and $\rho_H$ represent, respectively, the normal-resistivity and the Hall resistivity. 
A $\rho$ as a function of temperature for the film was measured in an applied field of ${\bf H}\ =$ 0 Oe 
and shown in FIG.\ref{Resistivity}. We note that the value of $\rho(T)$ was quite small 
at high temperature. The $\rho(T)$ behavior for pure TiO$_{2-\delta}$ film\cite{Yoon:2006} (solid line) 
is also shown in FIG.\ref{Resistivity} exhibiting a typical metal-insulator transition in the temperature 
range of 4 K and 300 K. In contrast to the $\rho$ for pure TiO$_{2-\delta}$, $\rho$ for the composite NG 
metallic Fe and TiO$_{2-\delta}$ film is a factor of 1000 lower and is constant for temperatures 
between 225 K 
and 400 K. The temperature variations are of the characteristic form expected from metal semiconductor interfaces. 
The $\rho$ value at 300 K was measured to be 183 $\mu \Omega - {\rm cm}$, and is about a factor of 20 larger 
than resistivity of pure Fe\cite{Weast:1982}. 
\begin{figure}[tp]
\centering
\includegraphics[width=0.43\textwidth]{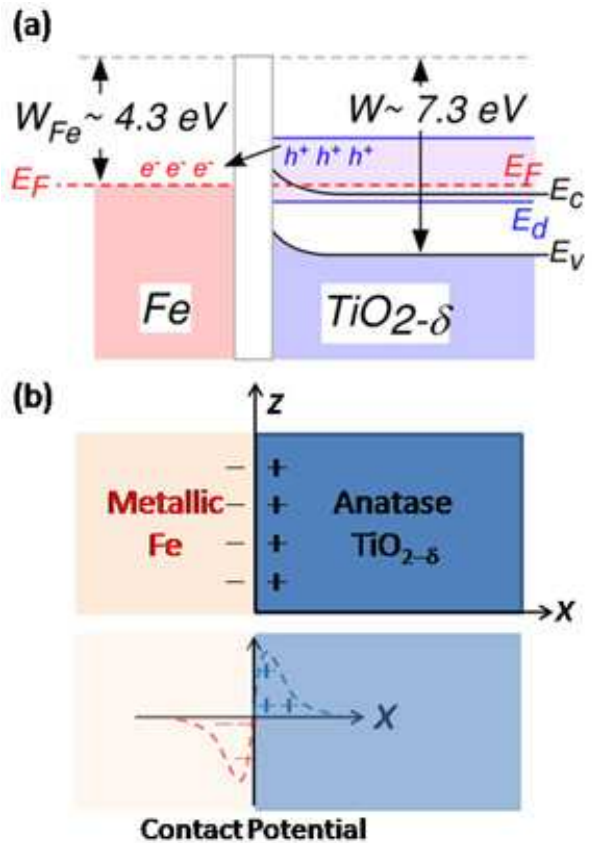}
\caption{(a) Energy band structure of nano granular metallic Fe and semiconducting TiO$_{2-\delta}$, 
where it is similar to metal oxide semiconductor. (b) Contact potential model.}
\label{MOS} 
\end{figure} 

Presence of these unique transport properties in the NG metallic Fe spheres embedded in TiO$_{2-\delta}$ films 
have been modeled as coexisting electronic structures of both semiconducting TiO$_{2-\delta}$ and 
NG metallic Fe. We have modeled the mechanism for transport in this composite film in a sketch shown 
in FIG.(\ref{MOS})a \cite{Streetman:1990,Knotek:1978,Zhang:1991}. A common chemical potential energy 
or Fermi energy in both TiO$_{2-\delta}$ and metallic Fe implies that the conduction band of TiO$_{2-\delta}$ 
is degenerate with the metallic Fe conduction band. Since the Fermi energy in metallic Fe is about 4.6 eV 
above the conduction energy level it implies that the energy band gap (3 eV) of TiO$_{2-\delta}$ 
is degenerate with the iron conduction band. This means that the donor levels in TiO$_{2-\delta}$ 
are also degenerate, and thereby, electrons hopping between Ti$^{2+}$ and Ti$^{3+}$ sites would find 
a conduit into the metallic conduction band leaving behind holes in TiO$_{2-\delta}$. 
Therefore, a small potential barrier must exist at the interface. This is illustrated schematically 
in FIG.\ref{MOS}a. As noted in FIG.\ref{TEM}, the NG metallic Fe sphere particles are isolated 
or disconnected, implying that conduction in the TiO$_{2-\delta}$ is via hole conduction. This can be 
also explained by contact potential at the interface between NG metallic Fe sphere and anatase TiO$_{2-\delta}$ 
as described in Landau Lifshitz\cite{Landau:1984}. In order to remove electron trough the surface 
of a metallic Fe, work must be done thermodynamically. According to the reference\cite{Landau:1984}, 
Poisson's equation for potential $\Phi (x)$ along the $x-axis$ normal to the interface between NG metallic 
Fe / TiO$_{2-\delta}$ implies  
\begin{eqnarray}
-\Phi^{\prime \prime } (x)=4\pi \rho (x),
\nonumber \\ 
\varpi  =\int_{-\infty }^\infty x\rho(x)dx=
-\frac{1}{4\pi } \int_{-\infty }^\infty x\Phi^{\prime \prime } (x)dx,
\nonumber\\
\Phi(\infty)- \Phi(- \infty)=V_c=4 \pi \varpi ,
\label{Contact-voltage}
\end{eqnarray}
wherein  $\varpi $ is defined to be dipole moment per unit contact area  
and $V_c$ is contact potential. The schematic of contact potential at the interface is shown 
in FIG.\ref{MOS}b.
\begin{figure}[tp]
\centering
\includegraphics[width=0.45\textwidth]{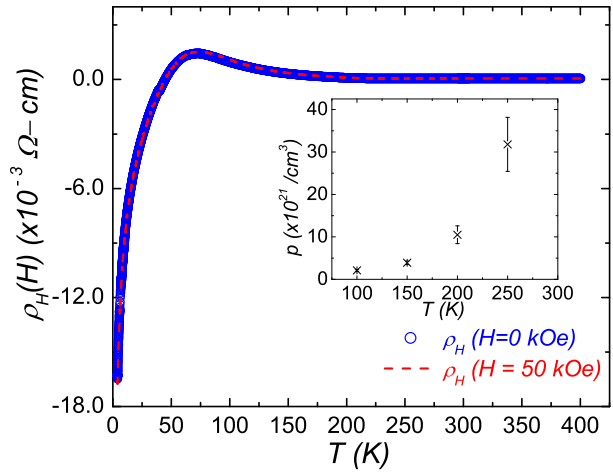}
\caption{(Hall resistivity, $\rho_H(H)$, versus temperature for H = 0 (circle symbol) and H = 50.0 kOe (dashed line) 
in the nano granular Fe in TiO$_{2-\delta}$ film.  Inset shows relation carrier density in function of temperature..}
\label{Hall-Resistivity} 
\end{figure} 

\begin{figure}[tp]
\centering
\includegraphics[width=0.45\textwidth]{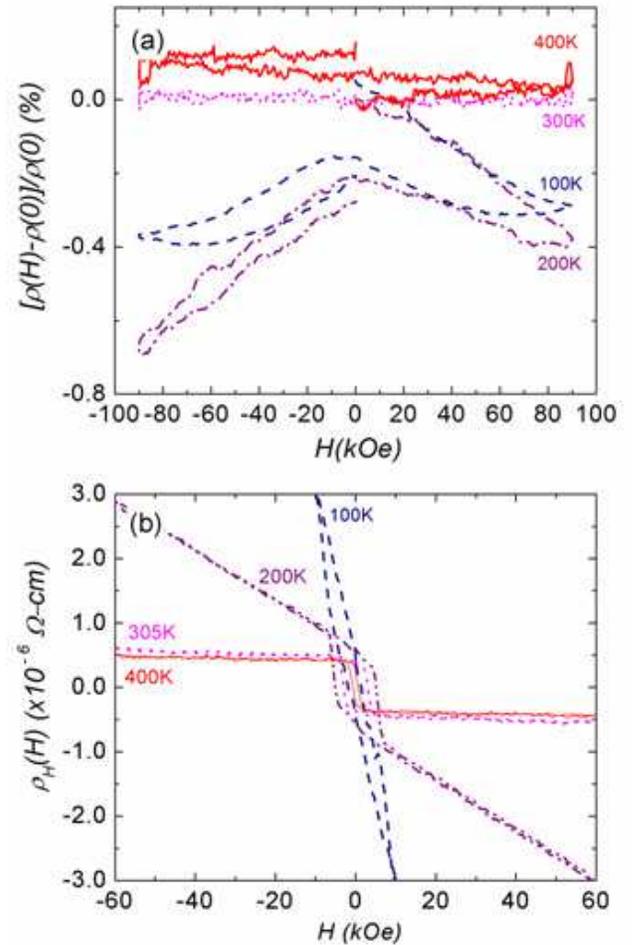}
\caption{(a) Magneto resistivity, $\rho(H)$ versus H in different temperatures. 
(b) Hall resistivity, $\rho_H(H)$ versus H in different temperatures.}
\label{MR-AHE-T} 
\end{figure} 
Electron conduction between metallic spheres may not be possible, see FIG.\ref{TEM}. In 
FIG.\ref{Hall-Resistivity}, 
Hall resistivity is plotted as a function of temperature that at high temperatures carriers are holes which also confirmed 
by Seebeck measurements. Furthermore, the number of carriers relative to pure TiO$_{2-\delta}$ has increased 
by factor 1,000 and the mass of holes is approximately about 10 times larger than the electron 
mass\cite{Yoon1:2007}. Also, in FIG.\ref{Hall-Resistivity} the carrier hole density {\it p} is plotted as a function 
of temperature. In order to calculate {\it p}, we employ
\begin{equation}
\rho_H \ = \ R_0B  +  4 \pi MR_S \ \ \ {\rm and} \ \ \ R_0 \ = \ \frac{1}{pec},
\label{AHE1}
\end{equation}
wherein $R_0$ and $R_S$ are the normal and anomalous Hall coefficients, respectively. 
In FIG.\ref{MR-AHE-T}, $\rho_H(H)$ scales linearly with magnetic field for temperatures below 300 K. 
The slope of $\rho_H(H) \ {\rm vs.} \ H$ gives rise to the normal Hall coefficient which corresponds 
to the first term in Eq.(\ref{AHE1}), where the hole carrier density, {\it p}, may be deduced. 
For temperatures well above 250 K, {\it p} is governed by the second term in Eq.(\ref{AHE1}), 
the anomalous Hall coefficient. Eq.(\ref{AHE1}) may re-written as\cite{Hurd:1972}
\begin{equation}
\rho_H \ = \frac{H}{pec} + \left(\frac{1}{pec} + R_S\right)(4 \pi M) 
\ \ {\rm where} \ \ R_S \gg R_0.
\label{AHE2}
\end{equation}
The anomalous Hall effect is dominant at high temperatures ($T\>>$ 250 K) and the second term 
in Eq.(\ref{AHE2}) is constant for saturation magnetization $M = M_S$.  

\begin{figure}[tp]
\centering
\includegraphics[width=0.42\textwidth]{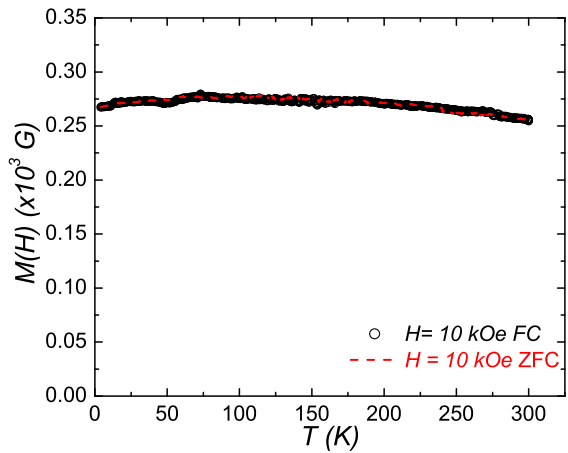}
\caption{Static magnetization, $M(H,T)$, versus temperature with $H \ =$ 10.0 kOe normal to the film plane.}
\label{MvsT} 
\end{figure}

The spontaneous magnetization, $M(H,T)$, was measured to be nearly constant as a function 
of temperature as shown in FIG.\ref{MvsT}. The measurement was performed with an external dc-magnetic field 
of 10.0 kOe applied normal to the film plane (out-of-plane measurement). The film can be fully saturated 
with a field of $\approx$ 6.6 kOe, since the coercive field was measured to be 2.0 kOe. Field cooled (FC) 
and zero-field-cooled (ZFC) $M(H,T)$ data in FIG.\ref{MvsT} shows no difference which implies 
that no spin glass effects are present in the films. There are two sources for magnetism 
in this composite structure: (1) ferromagnetism in TiO$_{2-\delta}$ and (2) ferromagnetism in metallic Fe spheres. 
The ferromagnetism in TiO$_{2-\delta}$ is due to double exchange as calculated by us in a previous 
calculation and it gives rise to a relatively small saturation magnetization at room temperature 
($\approx$ 400 Gauss). Based on the data presented in FIG.\ref{TEM}, we estimate a sphere density of 0.125 
to $0.25 \ \times \ 10^{18} \ {\rm spheres/cm}^3$ and sphere diameter between 10 and 15 nm. This implies that 
the loading factor of metallic Fe in the composite film is in the order of 0.05 to 0.25 which corresponds 

\begin{figure}[bp]
\centering
\includegraphics[width=0.48\textwidth]{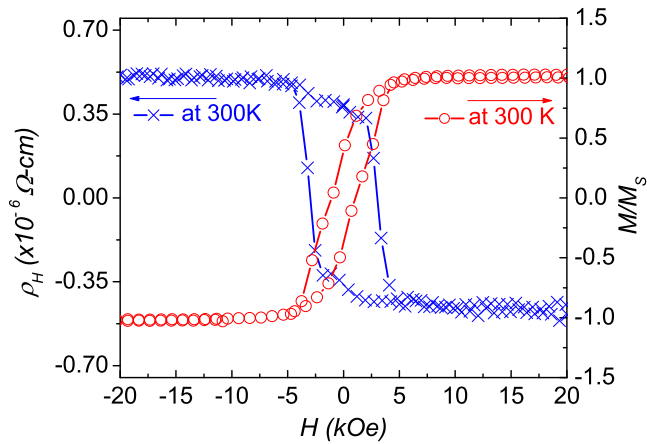}
\caption{Symbol (X) shows a typical anomalous Hall hysteresis loop, $\rho_H(H)$, at $T \ =$ 300 K 
with magnetic hysteresis loop (symbol (o)) at 300 K.}
\label{M-AHE-Hys} 
\end{figure}
to 1,100 to 5,100 Gauss for the saturation magnetization of the composite. This compares with a measured value 
of 3,100 Gauss. We have measured the transverse magnetoresistivity, $\rho(H)$, and Hall resistivity, $\rho_H(H)$, 
as a function of magnetic field sweeps between - 90 and + 90 kOe as shown in FIGS.\ref{MR-AHE-T}a and \ref{MR-AHE-T}b, 
and at different temperatures. In FIG.\ref{MR-AHE-T}a, the normal resistivity exhibits negative magnetoresistivity 
defined as, $\Delta f = (\rho(H)-\rho(0))/\rho(0)$ 
for temperatures $100\ K <T< 200\ K$.  $\Delta f$ is negligible for temperatures above 300 K. However, 
plots of $\rho_H(H)$ shown in FIG.\ref{MR-AHE-T}b, clearly demonstrate that saturation effects at temperatures 
near room temperature as a result of the hole carrier density increasing toward saturation levels, 
see FIG.\ref{Resistivity}. 
Notice that plots of $\rho_H(H)$ at $T \geq  200\ K$ exhibit clear hysteresis loop behavior similar 
to ferromagnetic hysteresis loops. Interestingly, saturation for both $\rho_H(H)$ and  $(M/M_S)$, 
normalized magnetization, occurred at exactly the same external magnetic field value of $\approx$ 6.6 kOe, 
see FIG.\ref{M-AHE-Hys}. This is no coincidence, since one requires a demagnetization factor of  
of $(4\pi /3)$ to magnetically saturate a sample of spherical shape in order to overcome 
a magnetic field of 6.8 kOe in Fe spheres at (say) room temperature. These data also show 
relatively smaller coercive field for the $(M/M_S)$ hysteresis loop than for the $\rho_H(H)$ 
hysteresis loop. The difference in hysteresis loops is due to the fact that the NG Fe spherical
samples give rise to dipole internal magnetic fields. In order to reverse the magnetization in each sphere it 
requires an external field to overcome this interactive internal field. Hence, the coercive field is approximately
constant with temperature as it scales with magnetization. In the Hall resistivity measurement the coercive field is
strongly temperature dependent, since there are two contributions to the resistivity measurement: 
(1) Normal Hall resistivity contribution which is not hysteretic with respect to $\bf{H}$ and (2) the anomalous Hall effects 
which is hysteretic with $\bf{H}$, since this effect is proportional to the magnetization. 
At high temperatures the AHE contribution dominates the resistivity measurement in contrast 
to low temperatures where the normal Hall contribution dominates. Thus, spin polarization as reflected 
in $\rho_H(H)$, see FIG.\ref{M-AHE-Hys}, correlates extremely well with the magnetic hysteresis loop implying that the magnetized 
Fe spheres are polarizing the carriers. 

\section{Discussion and Conclusions}
The plots of $\rho_H(H)$ at $T < 200\ K$ exhibit paramagnetic hysteresis behavior, see 
FIG.\ref{MR-AHE-T}b. 
This transition between paramagnetic and ferromagnetic $\rho_H(H)$ hysteresis behavior was also reported 
and correlated with hole carrier density related with RKKY theory in previous research\cite{Ohno:1998,Story:1986}. 
Their estimate of the threshold hole density for the carrier spin polarization transition 
was $p=3 \times 10^{20}/{\rm cm}^3$ from paramagnetic 
$p<3 \times 10^{20}/{\rm cm}^3$ to ferromagnetic $p\geq 3 \times 10^{20}/{\rm cm}^3$ 
behavior\cite{Ohno:1998,Hurd:1972,Story:1986}. Our estimate of hole carrier density of the  films was 
$p \approx 2.1 \times 10^{21}/{\rm cm}^3$ and $p \approx  \times 10^{22} /{\rm cm}^3$ at 
$T =$ 100 K and 200 K, 
respectively. The film exhibited strong polarization and hysteresis behavior as a function of applied field 
at 300 K. Since anomalous Hall effects may be observed in the presence of spontaneous 
magnetization\cite{Chien:1980,Karplus:1954}, we infer that about $\approx \ 10^{22} \ /{\rm cm}^3$ hole carriers 
are polarized by an applied field {\bf H}. This indicates that nearly all of the carriers were polarized 
by the spontaneous magnetization at $T \ =$ 300 K. According to FIG.\ref{M-AHE-Hys}, carrier polarization correlated very well 
with the magnetic hysteresis loop of NG Fe spheres. Also, FIG.\ref{M-AHE-Hys} indicates that the carrier polarization 
is not affected by external fields up to 3.0 kOe, which can be an advantage for memory device applications\cite{Prinz:1998,Wolf:2001}
However, if the shape of the particles embedded in the composite could be shaped into needles, it would imply 
spin polarization of carriers in external magnetic fields below 0.1 kOe which is ideally suited 
for spintronics applications.

In summary, magnetic and magneto-transport data for films of nano granular metallic Fe and oxygen defected 
TiO$_{2-\delta}$ are reported. The essence of this paper showed that conduction carriers of the films were 
strongly coupled to residual magnetic moments of metallic Fe grains in the nano composite structure. 
The dramatic reduction of normal resistivity ($\rho(T)$) of the films is a consequence of two factors: 
(1) oxygen defects in the TiO$_{2-\delta}$ host induced electron hopping; (2) electrons from the TiO$_{2-\delta}$ 
were introduced into the conduction band of Fe to create holes in TiO$_{2-\delta}$ similar 
to a Metal Oxide Semiconductor (MOS) structure. As a result the number of carriers increased at room temperature. 
The holes in TiO$_{2-\delta}$ were polarized due to the presence of ferromagnetic nano granular metallic Fe, 
where the carrier polarized density was measured to be near $\geq 3.0 \ \times 10^{22} \ /{\rm cm}^3$. 
Therefore, spintronics and spin dependent memory applications can be based upon the results presented here.
\section{Acknowledgment}
This research was supported by the National Science Foundation (DMR 0400676) and the Office of Naval Research 
(N00014-07-1-0701).

\end{document}